
\magnification 1200

\font\eightrm=cmr8
\font\eighti=cmmi8
\font\eightsy=cmsy8
\font\eightbf=cmbx8
\font\eighttt=cmtt8
\font\eightit=cmti8
\font\eightsl=cmsl8
\font\sixrm=cmr6
\font\sixi=cmmi6
\font\sixsy=cmsy6
\font\sixbf=cmbx6
\catcode`@11
\newskip\ttglue
\font\grrm=cmbx10 scaled 1200

\def\eightpoint{\def\rm{\fam0\eightrm}
\textfont0=\eightrm \scriptfont0=\sixrm \scriptscriptfont0=\fiverm
\textfont1=\eighti \scriptfont1=\sixi \scriptscriptfont1=\fivei
\textfont2=\eightsy \scriptfont2=\sixsy \scriptscriptfont2=\fivesy
\textfont3=\tenex \scriptfont3=\tenex \scriptscriptfont3=\tenex
\textfont\itfam=\eightit \def\it{\fam\itfam\eightit}
\textfont\slfam=\eightsl \def\sl{\fam\slfam\eightsl}
\textfont\ttfam=\eighttt \def\tt{\fam\ttfam\eighttt}
\textfont\bffam=\eightbf
\scriptfont\bffam=\sixbf
\scriptscriptfont\bffam=\fivebf \def\bf{\fam\bffam\eightbf}
\tt \ttglue=.5em plus.25em minus.15em
\normalbaselineskip=6pt
\setbox\strutbox=\hbox{\vrule height7pt width0pt depth2pt}
\let\sc=\sixrm \let\big=\eightbig \normalbaselines\rm}
\newinsert\footins
\def\newfoot#1{\let\@sf\empty
  \ifhmode\edef\@sf{\spacefactor\the\spacefactor}\fi
  #1\@sf\vfootnote{#1}}
\def\vfootnote#1{\insert\footins\bgroup\eightpoint
  \interlinepenalty\interfootnotelinepenalty
  \splittopskip\ht\strutbox 
  \splitmaxdepth\dp\strutbox \floatingpenalty\@MM
  \leftskip\z@skip \rightskip\z@skip
  \textindent{#1}\footstrut\futurelet\next\fo@t}
\def\fo@t{\ifcat\bgroup\noexpand\next \let\next\f@@t
  \else\let\next\f@t\fi \next}
\def\f@@t{\bgroup\aftergroup\@foot\let\next}
\def\f@t#1{#1\@foot}
\def\@foot{\strut\egroup}
\def\footstrut{\vbox to\splittopskip{}}
\skip\footins=\bigskipamount 
\count\footins=1000 
\dimen\footins=8in 

\def\ref#1{$^{#1}$}
\def\flex{\raise 6pt\hbox{$\leftrightarrow $}\! \! \! \! \! \! }

\newbox\bigstrutbox
\setbox\bigstrutbox=\hbox{\vrule height10pt depth5pt width0pt}
\def\bigstrut{\relax\ifmmode\copy\bigstrutbox\else\unhcopy\bigstrutbox\fi}
\def\refer[#1/#2]{ \item{#1} {{#2}} }
\def\rev<#1/#2/#3/#4>{{\it #1\/} {\bf#2}, {#3}({#4})}
\def\boxit#1{\vbox{\hrule\hbox{\vrule\kern3pt
\vbox{\kern3pt#1\kern3pt}\kern3pt\vrule}\hrule}}

\def\2figure#1#2#3#4{\vbox{ \hrule width#1truecm \hbox{\vrule height#2truecm
\hskip #1truecm
\vrule height#2truecm }\hrule width#1truecm \hbox{\vrule\vbox{\hsize #1truecm
\baselineskip=10pt
\noindent\strut#3}\vrule}\hrule width#1truecm
\hbox{\vrule\vbox{\hsize #1truecm
\baselineskip=10pt
\noindent\strut#4}\vrule}\hrule width#1truecm  }}
\def\3figure#1#2#3#4#5{\vbox{ \hrule width#1truecm \hbox{\vrule height#2truecm
\hskip #1truecm
\vrule height#2truecm }\hrule width#1truecm \hbox{\vrule\vbox{\hsize #1truecm
\baselineskip=10pt
\noindent\strut#3}\vrule}\hrule width#1truecm
 \hbox{\vrule\vbox{\hsize #1truecm
\baselineskip=10pt
\noindent\strut#4}\vrule}
\hrule width#1truecm \hbox{\vrule\vbox{\hsize #1truecm
\baselineskip=10pt
\noindent\strut#5}\vrule}\hrule width#1truecm  }}

\def\sqr#1#2{{\vcenter{\hrule height.#2pt
   \hbox{\vrule width.#2pt height#1pt \kern#1pt
    \vrule width.#2pt}
    \hrule height.#2pt}}}


\def\smin{\,\raise 0.06em \hbox{${\scriptstyle \in}$}\,}
\def\smsubset{\,\raise 0.06em \hbox{${\scriptstyle \subset}$}\,}

\def\Natural{\hbox{\hskip 1.5pt\hbox to 0pt{\hskip -2pt I\hss}N}}

\def\Rational{\hbox{\hbox to 0pt{\hskip 2.7pt \vrule height 6.5pt
                                  depth -0.2pt width 0.8pt \hss}Q}}
\def\Real{\hbox{\hskip 1.5pt\hbox to 0pt{\hskip -2pt I\hss}R}}
\def\Complex{\hbox{\hbox to 0pt{\hskip 2.7pt \vrule height 6.5pt
                                  depth -0.2pt width 0.8pt \hss}C}}

\baselineskip=14truept plus 2truept
\lineskip=2truept plus 1pt minus1truept
\lineskiplimit=1truept
\parskip=1truept plus 1truept
\hsize 6truein
\vsize 8.5truein
\centerline{\grrm On the amplitudes for non critical N=2 superstrings.}
\vskip 1cm
\centerline {E. Abdalla\ref 1 \ref 2 \newfoot{*}{permanent adress},
M.C.B. Abdalla$^3$, and D. Dalmazi$^2$}
\vskip .2cm

\centerline{$^1$ CERN Theory Division, 1211 Geneve 23, Switzerland}
\vskip .2cm

\centerline {$^2$Instituto de F\'\i sica, Univ. S\~ao Paulo, CP 20516,
S\~ao Paulo, Brazil}
\vskip .2cm

\centerline {$^3$Instituto de F\'\i sica Te\'orica, UNESP, Rua Pamplona 145,}
\vskip .2cm
\centerline {CEP 01405, S\~ao Paulo, Brazil}

\vskip 2cm
\centerline{ Abstract}

We compute correlation functions in $N=2$ non critical
superstrings on the sphere. Our calculations are restrained to the ($s=0$) bulk
amplitudes. We show that the four point function factorizes as a consequence of
the non-critical kinematics, but differently from the $N=0,1$ cases no extra
discrete state appears in the $\hat c\to 1^-$ limit.

\vfill\eject

\pageno=1
\footline={\hss\tenrm\folio\hss}

\noindent Critical N=2 strings have been recently considered by several
authors\ref{1,2,3}. Ooguri and Vafa\ref 1 computed explicitly scattering
amplitudes as, e.g., the four point function of the vertex operator
$$V(k)=\int d^4\theta d^2z e^{ik\cdot \overline X(z)+i\overline
k\cdot X(z)}$$
where $X$, $\overline X$,  are  the complex matter  superfields and the on
shell condition is given by $k\cdot \overline k = k_1\overline k_1 - k_2
\overline k_2 = 0$ . As it turns
out, the four point function vanishes for the critical $N=2$ superstring
theory\ref 1. It seems to be true that the higher functions do also vanish in
the critical theory.
This result has been obtained as a consequence of the kinematics
in 2+2 dimensional space-time. As a matter of fact,
this vanishing was already
expected, as argued by Ooguri and Vafa. Indeed, $N=2$ superstrings are
extremely simple string theories. Other string theories present an infinite
tower of particles, which should appear as bound states in the critical string
scattering amplitudes, as is the case of the Veneziano amplitude. However, in
the $N=2$ string, there is only a  massless, scalar particle; the only way of
obtaining consistency with the Veneziano amplitude and to avoid the infinite
tower of states seems to be  through
the vanishing of the amplitude. This is
actually what happens. There are several implications coming out of this
vanishing of higher point functions as discussed in [1].

Nevertheless, even critical $N=2$ strings are worthwhile studying. In fact,
$N=2$ theories are important objects in the study of integrable theories, and
string vacua\ref{4,5}. Moreover, there seems to be a strong relation between
self duality in four dimensions and integrability\ref{6}, a fact that has
extrapolated the barrier of dimensionality\ref{7}. Finally, we should mention
that there is a deep relation between integrable models and deformations of
conformally invariant theories\ref{8}, which although very interesting will not
concern us in the present work, but which might be important for $N=2$ in order
to understand the string vacuum.

Our present aim is to consider the non critical $N=2$ string theory. This might
be seen as a generalization of previous efforts to understand string theories
away from criticality\ref{9-18}. We will be actually concerned with a $N=2$
matter supermultiplet with $\hat c\le 1 \, (c=3\hat c)$ in a
super Coulomb gas representation conformally coupled to a $N=2$ superliouville
theory. However, as we shall see, the present case
contains a number of new technical difficulties, which in part is due to the
absence of the so called ``barrier" in the central charge. In fact, both
critical points coalesce, and the critical and non critical theories display a
unique amalgamation of their properties, enhancing the difficulties in
obtaining closed results.

\vskip1cm
\penalty-200

\noindent The appearance of Liouville theory as a byproduct of the
integration
over matter fields in a gravity background (as well as its supersymmetric
extension) is by now a very established issue\ref{19-21}. In the case of $N=2$
complex super Liouville ($S_{SL}$) theory interacting with a gravitational
field we consider the action
$$
\eqalign{
S =&  S_{SL} + S_M \cr
S_{SL}=&{1\over 4\pi}\!\!\int\!\! d^2w\hat E\left[\! \int\! d^4\theta\left(\Phi
\overline \Phi
-Q\hat Y(\Phi + \overline \Phi)\right) +\mu (\!\!\int \!\!d\theta^-d\overline
\theta^-e^{\overline \alpha\Phi} + \!\!\int \!\!d\theta^+d\overline
\theta^+e^{\alpha\overline \Phi})\right]\cr
S_M =& {1\over 4\pi}\!\!\int\!\! d^2w d^4\theta \hat E\left[ X\overline X +
 2i\alpha_0\hat Y(X +\overline X)\right] \cr}\eqno(1)
$$
The superfields $X\, ,\, \Phi \, (\overline X\, ,\, \overline \Phi) $ are
chiral (antichiral) and we have explicitly:
$$
\eqalign{
X (z,\overline z;\theta^-,\overline\theta^- )&=x (Z,\overline Z)+\psi_R
(Z,\overline Z) \theta^-+\psi_L(Z,\overline Z)\overline\theta^-+G(Z,\overline
Z)\theta^- \overline\theta^-\cr
\Phi (z,\overline z;\theta^-,\overline\theta^-)&=\varphi (Z,\overline Z)+\xi_R
(Z,\overline Z) \theta^-+\xi_L(Z,\overline Z)\overline\theta^-+F(Z,\overline
Z)\theta^- \overline\theta^-\cr}
\eqno(2)$$
where $(\theta^\pm)^\dagger =\overline \theta^\mp\, ,\, Z=z-\theta^+\theta^- $
and $ \overline Z=\overline z -\overline \theta^+\overline \theta^-$. The
quantity $\hat Y$ stands for the $N=2$ supercurvature superfield and $\hat
E$ for the superdeterminant of the superzweibein.

After setting $\mu=0$ in $S_{SL}$ we have the following expression for the last
component of the super energy momentum tensor (holomorphic part):
$$
\eqalign{
T&=T_{SL}+T_{M}\cr
T_{SL}&=-\colon \partial \overline \varphi\partial\varphi \colon
+{1\over 4}\colon \overline\xi_R
\partial\xi_R \colon + {1\over 4}\colon \xi_R\partial\overline\xi_R \colon
-{Q\over 2}\partial ^2(\varphi-\overline\varphi) \cr
T_M&=- \colon \partial\overline x \partial x \colon
+ {1\over 4}\colon \overline \psi_R\partial
\psi_R \colon + {1\over 4}\colon \psi_R\overline \partial \overline\psi_R
\colon + i\alpha_0\partial ^2(x+\overline x )\cr }\eqno(3)
$$

The first component of the super energy momentum tensor is given by the $U(1)$
current\ref{5} which generates the $U(1)$ symmetry of $N=2$
supersymmetric models. For the  (holomorphic) part of this currents we have:
$$\eqalign{
J&=J_{SL}+J_M\cr
J_{SL}&={1\over 4}\colon \overline\xi_R\xi_R\colon +{Q\over 2}
\partial(\varphi - \overline \varphi)\cr
J_{M}&={1\over 4}\colon \overline\psi_R\psi_R\colon +
i\alpha_0\partial( x - \overline x)\quad .
\cr }\eqno(4)$$
The propagators of the component fields can be read from the kinetic term of
(1):
$$
\eqalign{
\langle x(z)\overline x(w)\rangle&=\langle
\varphi(z)\overline \varphi(w)\rangle =\ln \vert z-w\vert ^{-2}\cr
\langle \psi_R(z)\overline \psi_R(w)\rangle &=\langle \xi_R (
z)\overline \xi_R (w) \rangle =2(z-w)^{-1}\cr
\langle \psi_L(\overline z)\overline \psi_L(\overline w)\rangle &=
\langle \xi_L (\overline z)\overline \xi_L (\overline w) \rangle
=2(\overline z-\overline w)^{-1}\cr}
\eqno(5)$$

Following [22] we fix $Q$ in (1) imposing the vanishing of the total
central charge
$$
\eqalign{
c_T& = c_{SL}+c_{M} + c_{ghosts}=0\quad ,\cr
c_{SL}&=3(1+2Q^2)\quad ,\cr
c_M&=3\widehat c \, ,\, \widehat c= 1-8\alpha_0^2\quad ,\cr
c_{ghosts}&=-6\quad ;\cr }\eqno(6)
$$

Thus we have:
$$Q=2\vert \alpha_0\vert \eqno(7)$$
where we chose $Q$ to be real; this corresponds to a choice of phases, as one
readily verifies.

The constants $\overline \alpha $ and $\alpha$ in eq. (1) can be fixed imposing
that the operators $e^{\overline \alpha \Phi}$ and $e^{\alpha \overline
\Phi}$ have dimension (1/2,1/2) (because of the double integration over the
Grasmann variables):
$$\eqalign{\Delta\left(e^{\overline\alpha\Phi}\right)&=-{\overline\alpha
Q\over 2}={1\over 2} \cr
\Delta\left(e^{\alpha\overline\Phi}\right)&=-{\alpha\overline Q\over
2}={1\over 2} \cr}\eqno(8)$$

$$\alpha = \overline \alpha = -{1\over\overline Q}=-{1\over Q}\eqno(9)$$

Note that the operator $e^{\alpha \Phi}\, (e^{\alpha \overline
\Phi})$ is chiral (antichiral) since it satisfies the chirality condition
$\Delta= +q\, (\Delta =-q)$, where the $U(1)$ charge $q$ of a $\varphi$ field
is defined as usual from the short distance expansion
$$
J(w)\varphi(z)={q\varphi\over w-z}+\cdots
\eqno(10)$$

It is easy to check that the values of $\alpha$ and $\overline \alpha$ in eq.
(9) assure vanishing $U(1)$ charge for the action $S_{SL}$ as required, with
the basic assignments $q(d\theta^+)=1/2=-q(d\theta^-)$. Therefore the
solution (9) is clearly a consistent one.

After having fixed the action, we must specify the vertex operators to
calculate correlation functions. For comparison with the critical\ref{1}  case
we shall be concentrated in vertex operators which are the analogous of the
tachyon vertex operators in the $N=0,1$ cases. However in the noncritical
theories the operators must be dressed by gravity. In the $N=2$ non-critical
case the vertex operator reads:
$$\eqalign{
&V(k,\overline k)=\int d^2zd^4\theta e^{i(k \overline
X+\overline k X) +\beta\overline \Phi + \overline \beta \Phi}
= \int d^2z e^{i(k\overline x +\overline k x) + \beta \overline \varphi +
\overline \beta \varphi}\cr
&\times  \left[ ik\partial\overline x +\beta \partial \overline \varphi
- i\overline k\partial x - \overline \beta \partial \varphi -
\overline k  k \overline\psi_R \psi_R + \overline \beta  \beta \overline
\xi _R\xi _R + i \overline \beta k \overline \psi_R\xi _R - i\overline k
\beta \psi_R\overline \xi _R\right]\cr
&\times  \left[ ik\overline \partial\overline x +\beta \overline
\partial \overline \varphi
- i\overline k\overline \partial x - \overline \beta \overline
\partial \varphi -
\overline k k \overline\psi_L \psi_L + \overline \beta\beta \overline
\xi _L\xi _L + i \overline \beta k \overline \psi_L\xi _L - i\overline k
\beta \psi_L\overline \xi_L \right]\quad .\cr}\eqno(11)$$
Notice that we have used  on shell expressions (with $\mu=0$) for the
superfields $X(\overline X)$:
$$
\eqalign{
X&=x(z,\overline z)+\psi_R(z)\theta^-+\psi _L(\overline z) \overline \theta ^-
- \partial x \theta ^+\theta^- - \overline \partial x \overline \theta
^+\overline \theta^-\cr
\overline X&=\overline x(z,\overline z)-\overline \psi_R(z)\theta^+ -
\overline \psi _L(\overline z) \overline \theta ^+
+ \partial \overline x \theta ^+\theta^- +\overline \partial \overline x
\overline \theta^+\overline \theta^-\cr }\eqno(12)
$$
and analogously for $\Phi\, ,\, \overline \Phi$. In equation (11) the complex
dressing $\beta$ is fixed as a function of the complex momentum $k$ by imposing
that the vertex $V(k,\overline k)$ be a dimensionless operator and its $U(1)$
charge
vanishes. This amounts respectively to:
$$
\eqalignno{
\Delta\left(e^{i(k \overline X+\overline k
X)+\beta\overline\phi+\overline\beta\phi }\right)\!& = \!{1\over 2}k(\overline
k-2\alpha_0) +{1\over 2}\overline k(k-2\alpha_0)\! -{1\over 2}\overline\beta
(\beta + Q)\! -{1\over 2} \beta(\overline \beta + Q)=0&\cr
q\left(e^{i(k \overline X+\overline k
X)+\beta\overline\phi+\overline\beta\phi }\right)\! & =\! {1\over 2}Q
(\beta - \overline\beta )+\alpha_0 (k-\overline k)=0\quad . &(13)\cr }
$$
The first equation fixes the real part of the dressing $\beta$ (up to a sign):
$$
\left( {\beta +\overline \beta \over 2}\right) _\pm = -{Q\over 2} \pm \left
\vert {k+\overline k\over 2} - \alpha_0\right \vert
\eqno(14)$$
In this paper we assume the positive sign solution which is equivalent
to positive
energy particles. The  equation  (13) fixes the imaginary part of $\beta$.

It will be convenient later
on to write the noncritical vertex in a form similar
to the critical one\ref{1}:
$$
\eqalign{
&V(k,\overline k) = \int  d^2z d^4\theta e^{i[(k\cdot \overline X) +
(\overline k\cdot X)]} = \int d^2z e^{i(k\cdot \overline x +\overline k
\cdot x)}\cr
&\times [ i k\cdot \partial \overline x - i \overline k \cdot \partial x -
(k\cdot \overline \psi _R)(\overline k\cdot \psi_R)] [ i k\cdot \overline
\partial \overline x - i \overline k \cdot \overline \partial x -
(k\cdot \overline \psi _L)(\overline k\cdot \psi_L)]\cr }\eqno(15)$$
where the scalar product is defined by $a \cdot b = a^1b^1 +a^2b^2$
and\newfoot{$^{*}$}{We use
the same symbol for the two component vector and for
its first component; there will be no room for confusion since the two
component vector will only appear inside scalar products, denoted by a
dot.} $k=(k,-i\beta)$,
$\overline k= (\overline k, -i\overline \beta)$, $X=(X,\varphi)$,
$\overline X=(\overline X,\overline \varphi)$, $ x=(x,\varphi)$,
$\overline x= (\overline x,\overline \varphi)$, $\psi_{R(L)}=(\psi_{R(L)},
\xi_{R(L)})$, $\overline \psi_{R(L)}=(\overline \psi_{R(L)},
\overline \xi_{R(L)})$. Notice that the second component of
the vector $\overline k$ is not the complex conjugate of the second component
of the vector $k$ in the non critical case

We are now  ready to compute correlation functions
$$
\langle V_{k_1}\cdots V_{k_n}\rangle = \prod_{i=1}^n\int d^2z_id^4\theta_i
\left\langle \prod_{i=1}^N e^{i(k_i\overline X_i + \overline k_i X_i) + \beta
_i\overline \phi_i + \overline \beta _i \phi_i }\right\rangle _{S_M+S_{SL}}
\eqno(16)$$

The first step is to integrate over the two zero modes $x_0^1,x_0^2$ of the
first component of the matter superfield ($x=x^1+ix^2$). The result gives
the conservation of the real and imaginary parts of the momenta $k$, both
are encoded in the following formula:
$$\sum _{i=1}^nk_i=2\alpha_0\eqno(17)$$

The next step is the integration over the Liouville zero modes\ref{9,10}
$\varphi_0^1,\varphi _0^2\, (\varphi \break =\varphi^1+i\varphi^2)$
this is more

delicate in the $N=2$ case. If we naively integrate over $\varphi_0^1$ and
$\varphi_0^2$ in eq.(16) we have a divergent result. It is not
clear\ref{23} how this divergence should be regularized. We opt for
making a Wick rotation in the zero modes such that $\varphi_0$ and $\overline
\varphi_0$ become real. After integrating over $\varphi_0,\overline \varphi_0$
we have:
$$
\eqalign{
{\cal A}_n & = {\mu^{s+\overline s}\over \alpha^2}
\Gamma (-s)\Gamma (-\overline s)
\int\prod_{i=1}^n d^2z_id^4\theta_i \Biggl\langle
\prod_{j=1}^ne^{i(k_j\overline X_j+\overline k_jX_j)+\beta_j{\overline
\phi}_j+\overline\beta_j\phi_j}\cr
&\times  \left(\int d^2\omega
d^2\overline\theta e^{\alpha{\overline\phi}}\right)^{\overline s}
\left(\int d^2 \eta d^2\theta e^{\alpha\phi}\right)^{
s}\Biggr\rangle\cr}\eqno(18)$$
where $s=-{1\over \alpha}(\sum \beta_i +Q)$, and $\overline s=-{1\over \alpha}
(\sum\overline \beta_i +\overline Q)$.
Although our regularization  is rather ad'hoc we believe that our
results for $s=0=\overline s$ bulk amplitudes are independent of this
procedure. From now on we only consider $s=\overline s=0$ bulk amplitudes:
$$
\sum _1^n\beta _i + Q =0 = \sum _1^n \overline \beta _i +  Q
\eqno(19)$$

We start by looking at the $n=3$ point function. After fixing the residual
$OSP(2,2)$ symmetry choosing $\theta_1^{(\pm)}=\theta_3^{(\pm)}=0$ and $
z_1=\infty\, ,\,, z_2=1\, ,\, z_3=0$ we have:
$$
\eqalign{
{\cal A}_3 = {(\ln \mu )^2\over \alpha ^2}&\Biggl\langle e^{ik_3\cdot x(0)}
e^{ik_2\cdot x(1)} [ik\cdot \partial \overline x - i\overline k\cdot
\partial x - (k\cdot
\overline \psi_R)( \overline k\cdot \psi_R)]\cr
&\times [ik\cdot \overline \partial \overline x - i\overline k\cdot
\overline \partial x - (k\cdot
\overline \psi_L)( \overline k\cdot \psi_L)]\Biggr\rangle_{S(\mu =0)}\cr}$$

$${\cal A}_3 = {(\ln \mu )^2\over \alpha^2} (c_{23})^2\eqno(20)$$
where
$$
\eqalign{
c_{ij}& = k_i\cdot \overline k_j - \overline k_i \cdot k_j \cr
& = k_i \overline k_j - \overline k_i k_j - \beta _i\overline \beta _j
+\overline \beta _i\beta _j \cr }
\eqno(21)$$
and $(\ln \mu)^2=$ finite part of $\left( \lim\limits_{\overline s,s\to 0
}\mu^{s + \overline s}\Gamma(-s)\Gamma(-\overline s)\right)$.
In order to rewrite
${\cal A}_3$ in a more sugestive form we need some kinematics. First of all it
is easy to show (using (13),(17) and (19)) that
$$
\sum _{j=1}^n c_{ij} = 2\alpha_0 (k_i - \overline k_i ) + (\beta _i - \overline
\beta _i)Q =0
\eqno(22)$$

The vanishing of $\sum _{j=1}^nc_{ij}$ holds in the critical\ref{1} case as a
consequence of the momentum conservation and the on shell condition
$k\cdot\overline k=k_1\overline k_1-k_2\overline k_2=0$. It is remarkable that
(22) holds also in the non critical case as a consequence of the zero $U(1)$
current condition (13).

We assume\newfoot{$^{*}$}{The calculations for $\alpha_0>0$ are completely
analogous.} from now on that $\alpha_0<0$ in this case we
have from (13) and (14)
$$
\beta (k) = \cases {k & , if $ \, \, \Re\!e  k ={k+\overline k \over 2}
>\alpha_0 $ \cr 2\alpha_0-\overline k  & , if $\,\, \Re\!e k <\alpha_0$\cr}
\eqno(23)$$
therefore
$$
k\cdot \overline k =k \overline k -\beta \overline \beta = \cases { 0 & ,
 if $\, \, \Re\!e  k > \alpha_0 $\cr
4\alpha_0 ( \Re\!e  k - \alpha _0) & ,  if $ \, \, \Re\!e  k < \alpha_0 $\cr}
\eqno(24)$$

Using all these kinematic relations we may write ${\cal A}^3 $ from (17) in the
region $\Re\!e k_2\, ,\, \Re\!e k_3< \alpha_0\, ,\,
\Re\!e k_1>\alpha_0$ in a factorized
form:
$$
{\cal A}_3 ={(\ln \mu )^2\over \alpha^2} {(\Im\!m k_1)^2\over
\alpha_0^2} (k_2\cdot \overline k_2)(k_3\cdot \overline k_3)
\eqno(25)$$
where $ \Im\!m  k =(-i) {(k-\overline k)\over 2}$.
The amplitude vanishes for any other kinematic region, where at least two
momenta satisfy $Re k>\alpha_0$ and are therefore ``on shell" $(k\cdot
\overline k=0)$ in the critical sense. It should be stressed that the amplitude
${\cal A}_3$ in the critical case\ref{1} has the same form (17) but it cannot
be written in a factorized form as in (25).

Now we calculate the four-point amplitude ${\cal A}_4$. Fixing
$\theta_1^{(\pm)}=\theta_4^{(\pm)}=0$ and $z_1=\infty\, ,\, z_2=1\, ,\, z_3=z\,
,\, z_4=0$ we have , after using (22), basically the same expression as in the
critical case:
$$
{\cal A}_4=\!{(\ln \mu )^2\over 16\alpha^2}
\!\!\int\!\! d^2z \vert z \vert ^{-s} \vert
1-z\vert ^{-t}\!\left( {t(t+2)\over (1-z)^2} \! + \!{4c_{12}c_{34}\over z}
\!+ \!
{4c_{23}c_{41}\over 1-z}\right) \times ({\hbox { ``h. c." }})
\eqno(26)$$
where\newfoot{$^{**}$}{Our definition of $s$ and $t$ correspond to twice of
ref.[1]
because their propagator correspond to half of ours (see (5))} $ s=-2s_{34}\,
,\, t=-2s_{23}$ and $s_{ij}=k_i\cdot \overline k_j+\overline
k_i\cdot k_j$. The hermitian conjugated part above corresponds to the previous
term inside the brackets with $\overline z$ instead of $z$. Note that it is not
really the hermitian  conjugated expression since $\overline k_i$ is not the
complex conjugated of $k_i$ as we stressed before.

After performing the integrals in (26) using formulas of ref.[1] and making
algebraic manipulations which are consequence of kinematic relations common to
the critical and noncritical cases we have:
$$
{\cal A}_4 = - {\pi (\ln \mu )^2\over \alpha^2}F^2 \Delta (s_{34})\Delta
(s_{14})\Delta (s_{24})
\eqno(27)$$
where
$$
F = [ (k_1\cdot \overline k_2)(k_2\cdot \overline k_3 )( k_3 \cdot \overline
k_1) + {\hbox { ``h. c."}} ]
\eqno(28)$$
and $\Delta (x)=\Gamma(x)/\Gamma(1-x)$.

The expression (27) is essentially the same one derived in critical case, the
difference now comes from the fact that after fixing the kinematic region
$\Re\!e k_1$, $\Re\!e k_2$, $\Re\!e k_3 <\alpha_0$, $\Re\!e k_4>\alpha_0$
we have, after a long algebra, a very simple expression for $F$:
$$
F=- {\left( (\Im\!m  k_4)^2 + \alpha_0^2\right)
 \over 4\alpha_0^2}\prod_{i=1}^3
(k_i\cdot \overline k_i).
\eqno(29)$$
In the above kinematic region
 we also obtain $s_{i4}=-(k_i\cdot \overline k_i)$,
therefore we can finally write ${\cal A}_4$ in a factorized form:
$$
{\cal A}_4 = {\pi (\ln \mu )^2\over 16\alpha^2} {\left( (\Im\!m  k_4)^2
+\alpha_0^2\right)^2\over \alpha_0^4 }\prod _1^3
\Delta (1-k_i\cdot \overline k_i).
\eqno(30)$$
As in the case of the 3-point function (${\cal A}_3$), it can be shown that
${\cal A}_4$ vanishes in any kinematic region where at least two momenta
satisfy $\Re\!e k> \alpha_0$.

In the critical limit $\hat c \to 1\, (\alpha_0\to 0)$ we have $\Delta
(1-k_i\overline k_i)\sim \alpha_0$ thus,
$$
{\cal A}_4 (\alpha_0\to 0) \sim {(\Im\!m  k_4 )^4 \over \alpha_0\alpha^2 }
(\ln \mu)^2. \eqno(31)$$
If we absorb the factor $1/\alpha^2\, (=4\alpha_0^2)$ in the measure
of the path integral the amplitude ${\cal A}_4$ diverges like $1/\alpha_0$ ,
otherwise  the amplitude vanishes ${\cal A}_4\sim \alpha_0$ as in the critical
case\ref{1}. It should be noticed, however,
 that the factor $1/\alpha^2$ (which
comes from the double zero mode integrals) must be absent in ${\cal A}_3$ (see
(17)) if we want  to obtain the critical result in the $\alpha_0\to 0$ limit,
otherwise we would have a vanishing 3-point coupling. Thus we can not obtain
both ${\cal A}_3 $ and ${\cal A}_4$ of the critical case in the $\alpha_0 \to
0$ limit  (see conclusion).

For $\hat c< 1$ the
interesting models are the minimal ones and
 in these cases it is easy to show
that the functions $\Delta(1-k_i\cdot \overline k_i)$ have no poles
(or zeroes) thus,
they can be simply absorbed in
 the definition of the vertices $V(k,\overline k)$
exactly as in the $N=0,1$ cases. Actually the factor $\Delta (1-k_i\overline
\cdot k_i)= \Delta(1+\beta\overline \beta-k\overline k)$ corresponds to the
factors\ref{12} $\Delta \left( {1\over 2} (1+\beta^2-k^2)\right)$ and\ref{13}
$\Delta\left({1\over 2}(\beta^2 - k^2)\right)$ of the $N=1$ and $N=0$
cases respectively, in the sense that all these factors become $\Delta(1)=0$
for configurations with zero
 energy $(E= \Re\!e  \beta +{Q\over 2}=0)$ which sho
ws the
decoupling of such states\ref{24}.

\vskip .3cm
\centerline {\bf Conclusions}
\vskip .2cm
We have calculated the three and four point $(s=0)$ bulk amplitude, in a
non-critical $N=2$ superstring
 consisting of a $N=2$ matter supermultiplet with
central charge $\hat c\le 1\, (c=3\hat c)$ conformally coupled to an $N=2$
super Liouville theory.
 We have shown that both amplitudes may be written ,in a
certain kinematic region,
in a factorized form. For other kinematic regions the
amplitudes vanish. Moreover after a suitable renormalization of the measure of
the path integral we recover the result for the three-point amplitude of the
critical case\ref{1} in the limit $\alpha_0 \to 0\, (\hat c \to 1^-)$. There
is, however, an amazing difference with respect to the critical string when we
look at the four-point amplitude in the same limit, namely, we get a divergent
result $({\cal A}_4 \sim 1/\alpha_0)$ rather than a vanishing one\ref{1}. The
difference with the critical case can be explained as follows, the prefactor
$F$ in (29), which vanishes identically in the critical case\ref{1}, goes to
zero in the critical limit $(F\sim \alpha_0)$ but its vanishing is supressed
by the poles of the functions $\Delta (-k_i\cdot \overline k_i)\, (
\Delta (-k_i\cdot \overline k_i)  \sim 1/\alpha_0)$ which usually correspond
to intermediate states\newfoot{$^*$}{Since $\alpha_0$ works like  an infrared
cut-off in non-critical calculations\ref{13} we might even speculate that such
poles correspond to the exchange of the massless particle present in the
critical theory. The non vanishing contribution of those poles in the non
critical case might be
attributed to the non trivial kinematics.}, such poles cancel in the
critical case but they show up in the non critical one as a direct consequence
of the non analytic structure of
the dispersion relation (14) which permited to
fix completelly, in a certain
 kinematic region, the real part of the momenta of
one of the scattered particles (in our case $\, \Re\!e k_4=-\alpha_0$).
This means
that there may not exist a smooth limit from the non critical to the critical
case. We should also mention that it is possible to redefine the
vertices $V(k,\overline k)$ and the path integral measure by powers of
$\alpha_0$ such that the amplitudes become finite in the $\alpha_0  \to 0$
limit. In order to conclude the discussion about the $\hat c \to 1^-$ limit
we remark that
whatever is the correct interpretation of ${\cal A}_4$ we do not have extra
discrete states as in the $N=0,1$ cases
 and this is expected since the spectrum
of critical $N=2$ string is finite.

For $\hat c <1$ the effect of the functions $\Delta (-k_i\cdot \overline k_i)$
is as mild as in the $N=0,1$ cases and the $s=0$ amplitudes are basically
given by the factor $(\ln \mu)^2$.

Several aspects of our results are still unclear and a more carefull analysis
is needed, wich would imply the calculation of higher point amplitudes as well
as $s\ne 0$ correlators, this is under progress.

Acknowledgements: the authors would like to thank discussions with profs. A.
Schwimmer, G. Sotkov, and M. Stanishkov. This work was partially supported by
FAPESP (S\~ao Paulo) and CNPq (Brazil).

\vskip .2truecm
\penalty-1000

\centerline{\bf  Reference}

\vskip .1truecm
\refer[[1]/H. Ooguri, C. Vafa, Nucl. Phys. {\bf B361} (1991)469, {\bf B367}
(1991)83.]

\refer[[2]/A. Giveon and A. Shapere, Cornel and Princeton preprints
CLNS-92/1139, IASSNS-HEP-92/14.]

\refer[[3]/M. Li, Univ. California, Santa Barbara, UCSBTH-92/14]

\refer[[4]/C. Vafa, at the Simp. on Fields, Strings and
Quantum gravity, Beijing, 1989, and Summer School in High Energy Physics and
Cosmology, ICTP, 1989, 1990.]

\refer[[5]/P. Fendley, S. D. Mathur, C. Vafa and N. P. Warner, Phys. Lett. {\bf
B243} (1990)257.]

\refer[[6]/M. K. Prasad, A. Sinha, L. L. Chau Wang, Phys. Rev. Lett. {\bf 43}
(1979)750. ]

\refer[[7]/E. Witten Nucl. Phys {\bf B266} (1986)255; J. Avan, H. J. de Vega
and J. M. Maillet Phys. Lett.
 {\bf 171B} (1986)255;   E. Abdalla, M. Forger and
M. Jacques Nucl. Phys. {\bf B307} (1988)198.]

\refer[[8]/V. A. Fateev, Int. J. Mod. Phys. {\bf A12} (1991)2109; P. Fendley
and K. Intriligator HUPT-91/A067.]

\refer[[9]/A. Gupta, S. P. Trivedi and M. B. Wise, Nucl. Phys. {\bf B340}
(1990)475.]

\refer[[10]/M. Goulian and M. Li, Phys. Rev. Lett. {\bf 66} (1991)2051.]

\refer[[11]/Vl. S. Dotsenko Mod. Phys. Lett {\bf A6 } (1991)3601.]

\refer[[12]/E. Abdalla, M. C. B. Abdalla, D. Dalmazi, K. Harada, Phys. Rev.
Lett. {\bf 68 }  (1992) 1641; Int. J. Mod. Phys. {\bf A } to appear.]

\refer[[13]/ P. di Francesco and D. Kutasov, Nucl Phys. {\bf B375}(1992)119.]

\refer[[14]/I. Klebanov, Priceton Preprint PUPT-1271, 1991.]

\refer[[15]/K. Aoki, E. d'Hoker, Mod. Phys. Lett. {\bf A7} (1992) 235,333.]

\refer[[16]/L. Alvarez-Gaum\'e, Ph. Zaugg, Phys. Lett. {\bf B273} (1991)81]

\refer[[17]/L. Alvarez-Gaum\'e and J. L. Ma\~nez, Mod. Phys. Lett. {\bf A6}
(1991)2039;  L. Alvarez-Gaum\'e, H. Itoyama, J. L. Ma\~nez and A. Zadra, CERN
prep, 1992]

\refer[[18]/G. Moore, N. Seiberg, M. Staudacher, Nucl. Phys. {\bf
B362}(1991)665]

\refer[[19]/A. M. Polyakov Phys. Lett. {\bf B103} (1981)207.]

\refer[[20]/A. M. Polyakov Phys. Lett. {\bf B103} (1981)211.]

\refer[[21]/O. Alvarez Nucl. Phys. {\bf B216} (1986)125;{\it ibid } {\bf
B238}(1984) 61; E. Abdalla, M. C. B. Abdalla and
K. D. Rothe, Non perturbative methods in 2 dimensional quantum, World
Scientific Publising, 1991.]

\refer[[22]/J. Distler, Z.Hlousek, H. Kawai, Int. J. Mod. Phys. {\bf A5}
(1990)391.]

\refer[[23]/N. Seiberg and D. Kutasov, Phys. Lett. {\bf B251}(1990)67.]

\refer[[24]/N. Seiberg, Lecture at 1990 Yukawa Int. Sem. Common Trends in Math.
and Quantum Field Theory, and Cargese meeting Random Surfaces, Quantum Gravity
and Strings, May 27, June 2, 1990; J.
Polchinski, Strings '90 Conference, College Station, TX, Mar 12-17, 1990, Nucl.
 Phys. {\bf B357}(1991)241.]

\end